\documentclass[conference]{IEEEtran}
\usepackage[dvips,colorlinks,bookmarksopen,bookmarksnumbered,citecolor=black,urlcolor=black]{hyperref}
\usepackage{moreverb}
\usepackage[dvips]{graphicx}
\usepackage{amsfonts}
\usepackage{amssymb}
\usepackage{amsbsy}
\usepackage{amsmath}
\usepackage{amsthm}
\usepackage{tikz}
\ifCLASSINFOpdf
\else
\fi
\begin{document}
\title{Analysis of Coverage Region for\\ MIMO Relay Channel}

\author{\IEEEauthorblockN{Alireza Alizadeh}
\IEEEauthorblockA{Department of Electrical Engineering\\
Ferdowsi University of Mashhad\\
Mashhad, Iran\\
Email: alr.alizadeh@gmail.com}
\and
\IEEEauthorblockN{Ghosheh Abed Hodtani}
\IEEEauthorblockA{Department of Electrical Engineering\\
Ferdowsi University of Mashhad\\
Mashhad, Iran\\
Email: ghodtani@gmail.com}}


%Telephone: (800) 555--1212, Fax: (888) 555--1212}


\maketitle

\begin{abstract}
%\boldmath
In this paper we investigate the optimal relay location in the sense of maximizing suitably defined coverage region for MIMO relay channel. We consider the general Rayleigh fading case and assume that the channel state information is only available at the receivers (CSIR), which is an important practical case in applications such as cooperative vehicular communications. In order to overcome the mathematical difficulty regarding determination of the optimal relay location, we provide two analytical solutions, and show that it is possible to determine the optimal relay location (for a desired transmission rate) at which the coverage region is maximum. Monte Carlo simulations confirm the validity of the analytical results. Numerical results indicate that using multiple antennas increases coverage region for a fixed transmission rate, and also increases the transmission rate linearly for a fixed coverage.
\end{abstract}


% \fi
%\IEEEpeerreviewmaketitle

\section{Introduction}
%\IEEEPARstart{T}{he}
The relay channel is the most basic structural unit in wireless networks and relaying strategy can increase channel throughput, coverage region, facilitate information transmission and realizes some of the gains of multiple-antenna systems by single-antenna terminals. The relay channel, since its introduction by van der Muelen~\cite{VM}, has been extensively studied~\cite{CE}-\cite{HA2}. In their seminal work, Cover and El Gamal~\cite{CE} presented a capacity upper bound and achievability strategies for the relay channel. Although the channel capacity is still unknown, the authors in~\cite{HA2} unified most of known capacity theorems into one capacity theorem, which potentially may be applicable to a more general class of relay channels.

Using multiple antennas can be considered as an effective technique to combat fading which also can result in a significant increase in channel throughput. Nonetheless, the capacity gain obtained from this technique heavily depends on the amount of instantaneous CSI available at the receivers and transmitters~\cite{GJJV}. Considering different scenarios of CSI, the capacity for point-to-point MIMO channel in~\cite{T} and for MIMO relay channel in~\cite{WZH}-\cite{NF} have been analyzed.

%\subsection{Our Work in Relation to the Previous Work}
The existing results for MIMO relay channels mainly focus on maximizing capacity bounds for fixed locations of the channel nodes. However, in most wireless networks, the optimal location of the relay and coverage region is of practical interest due to the mobility of the destination (e.g. a mobile station in a cellular network). The authors in~\cite{ABC} have investigated single-antenna Gaussian relay channel with the new objective of maximizing coverage for a desired transmission rate, and by considering decode-and-forward (DF) and compress-and-forward (CF) strategies for the relay channel, optimized the relay location as a design parameter. In this paper, we define the concept of coverage and investigate the coverage region in a more general and practical case, i.e., MIMO fading relay channel with only CSIR. More precisely, our goal is to determine the optimal relay location (for a desired transmission rate) at which the coverage region is maximum. Thus, considering the coverage definition and the necessary condition for applying DF strategy, we express the desired transmission rate in terms of the optimal relay location and derive two expressions with the help of which we can determine the optimal relay location.

%\subsection{Paper Organization}
The rest of the paper is organized as follows. We start with the channel model in section II. In section III, by defining the channel path-loss coefficients, we review the capacity bounds of MIMO relay channel in an appropriate form for the case when the channel entries are i.i.d. Rayleigh fading and only CSIR is available. The concept of coverage and evaluation of optimal relay location are provided in section IV. Numerical results are presented in section V, and section VI contains our conclusion.
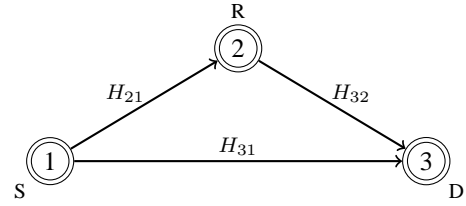
\begin{figure}[t]
	\centering
\begin{tikzpicture}
\node at (2.5,2) {\footnotesize R};
\node at (-.4,-.4) {\footnotesize S};
\node at (5.4,-.4) {\footnotesize D};
\node at (2.5,.2) {\footnotesize$H_{31}$};
\node at (1,0.9) {\footnotesize$H_{21}$};
\node at (4,0.9) {\footnotesize$H_{32}$};
\draw (0,0) circle [radius=.25];
\draw (2.5,1.5) circle [radius=.25];
\draw (5,0) circle [radius=.25];
\node[circle,draw]
(A) at (0,0) {1};
\node[circle,draw]
(B) at (2.5,1.5) {2};
\node[circle,draw]
(C) at (5,0) {3};
\thicklines
\draw[->,thick] (A) -- (B);
\draw[->,thick] (B) -- (C);
\draw[->,thick] (A) -- (C);
\end{tikzpicture}
		\caption{MIMO relay channel.}
	\label{Fig1}
\end{figure}

%\subsection{Notations}
Throughout this paper, scalars are represented by lowercase letters, vectors are denoted by lowercase boldface letters and uppercase boldface letters are used for matrices. Superscripts \emph{T} and \emph{H} denote the transpose and conjugate transpose, respectively. The operator $\mathcal{E}[.]$ stands for the expectation and $\log(.)$ denotes base-2 logarithm. $\boldsymbol{I}_M$ is the $M\times M$ identity matrix, and we use $\mathrm{tr}(\boldsymbol{A})$ and $\det(\boldsymbol{A})$ to denote the trace and the determinant of the matrix $\boldsymbol{A}$, respectively. The distribution of a circularly symmetric complex Gaussian vector $\boldsymbol{x}$ with mean $\boldsymbol{\mu}$ and covariance matrix $\boldsymbol{Q}$ is denoted by $\boldsymbol{x}\sim\mathcal{CN}(\boldsymbol{\mu},\boldsymbol{Q})$, and $z\sim\mathcal{CN}(0,1)$ is a circularly symmetric complex Gaussian random variable, where its real and imaginary parts are zero mean i.i.d. Gaussian random variables, each with variance 1/2, i.e., $\mathcal{N}(0,1/2)$.
\begin{figure}[t]
	\centering
	
\begin{tikzpicture}[scale=0.75]
\draw [help lines,dashed] (0,0) grid (8,3);
\draw [thick, <->] (0,3) -- (0,0) -- (8,0);
\node at (8.3,-.05){$x$};
\node at (0,3.3){$y$};
\node at (.4,.4){\footnotesize S};
\node at (5,.4){\footnotesize R};
\node at (6,3.4){\footnotesize D};
\draw [fill] (0,0) circle [radius=.1];
\node at (0,-.4){\footnotesize $(0,0)$};
\draw [fill] (5,0) circle [radius=.1];
\node at (5,-.4){\footnotesize $(u_2,0)$};
\draw [fill] (6,3) circle [radius=.1];
\node at (6,2.6){\footnotesize $(u_3,v_3)$};
%\draw [blue] (0,0)--(6,3);
%\draw [blue] (5,0)--(6,3);
\end{tikzpicture}
		\caption{Network geometry.}
	\label{Fig2}
\end{figure}
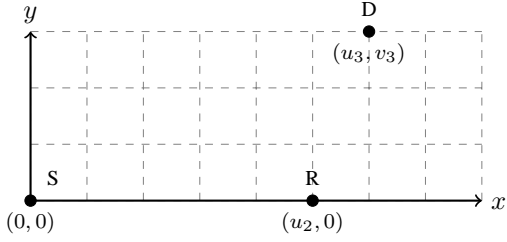
\section{Channel Model}
\label{sec2}
Consider the MIMO relay channel in Fig.~\ref{Fig1}. The source transmits a message $\boldsymbol{x}_1$ over the channel. Let us consider $\boldsymbol{y}_2$ as the received signal at the relay; based on prior received signals, the relay then transmits a message $\boldsymbol{x}_2$ that is intended to facilitate the transmission between the source and the destination. We assume the source has $M_1$ transmit antennas, and the destination has $N_3$ receive antennas. Also, suppose that the relay is equipped with $M_2$ and $N_2$ antennas for transmitting and receiving, respectively. Thus, the received signals at the destination and the relay can be expressed as
\begin{equation}
\label{eq1}
\boldsymbol{y}_3=\boldsymbol{H}_{31}\boldsymbol{x}_1+\boldsymbol{H}_{32}\boldsymbol{x}_2+\boldsymbol{n}_3
\end{equation}
\begin{equation}
\label{eq2}
\boldsymbol{y}_2 = \boldsymbol{H}_{21} \boldsymbol{x}_1 + \boldsymbol{n}_2
\end{equation}
where
\begin{itemize}
\item $\boldsymbol{x_1}$, $\boldsymbol{x_2}$ are $M_1\times 1$ and $M_2\times 1$ transmitted signals of the source and the relay, respectively; and have zero mean, i.e., $\mathcal{E}[\boldsymbol{x_1}]=\boldsymbol{0}$, $\mathcal{E}[\boldsymbol{x_2}]=\boldsymbol{0}$;
\item $\boldsymbol{y_3}$, $\boldsymbol{y_2}$ are $N_3\times 1$ and $N_2\times 1$ received signals at the destination and the relay, respectively;
\item $\boldsymbol{H}_{31}$, $\boldsymbol{H}_{21}$, $\boldsymbol{H}_{32}$ are $N_3\times M_1$, $N_2\times M_1$, and $N_3\times M_2$ channel gain matrices as shown in Fig.~\ref{Fig1}, and modeled as independent (flat) fading processes;
\item $\boldsymbol{n_3}$, $\boldsymbol{n_2}$ are respective independent $N_3\times 1$ and $N_2\times 1$ zero-mean circularly symmetric complex Gaussian noise vectors at the destination and the relay, with distribution $\mathcal{CN}(\boldsymbol{0},\boldsymbol{I}_{N_3})$ and $\mathcal{CN}(\boldsymbol{0},\boldsymbol{I}_{N_2})$, which are independent of the transmit signals.
\end{itemize}
Considering $M=M_1+M_2$, the $M\times M$ joint transmit covariance matrix of the zero-mean source and relay transmit signals can be described by
\begin{equation}
\label{eq3}
\boldsymbol{Q}\triangleq
\begin{bmatrix}
\boldsymbol{Q_{11}}& \boldsymbol{Q_{12}}\\
\boldsymbol{Q_{21}}& \boldsymbol{Q_{22}}
\end{bmatrix}
\end{equation}
where $\boldsymbol{Q}_{ij}=\mathcal{E}[\boldsymbol{x}_i \boldsymbol{x}_j^{H}], i,j=1,2$ is the covariance matrix between the input signals $\boldsymbol{x}_i$ and $\boldsymbol{x}_j$. Note that $\boldsymbol{Q}$ is a Hermitian matrix~\cite{T}. Throughout this paper, we suppose that the relay works in full-duplex mode (the relay can transmit and receive in the same frequency band at the same time), because it can be considered as a performance upper bound for half-duplex systems~\cite{NF}.

Now consider the network geometry depicted in Fig.~\ref{Fig2}. In this configuration the source and the relay are located at $d_1=(0,0)$ and $d_2=(u_2,0)$, respectively, and the destination is located at $d_3=(u_3,v_3)$. Also, we suppose that $\eta$ is the distance-based path-loss power attenuation exponent. Thus, the channel matrices can be represented as follows
\begin{equation}
\label{eq4}
\boldsymbol{H}_{31}=a.\boldsymbol{H}_{w31},~~\boldsymbol{H}_{21}=b.\boldsymbol{H}_{w21}, ~~\boldsymbol{H}_{32}=c.\boldsymbol{H}_{w32}
\end{equation}
where the channel path-loss coefficients defined as:
${a\triangleq(u_{3}^{2}+v_{3}^{2})^{(-\eta/4)}}$, ${b\triangleq u_2^{-\eta/2}}$, ${c\triangleq((u_2-u_3)^2+v_3^2)^{-\eta/4}}$, the entries of $\boldsymbol{H}_{w31}$, $\boldsymbol{H}_{w21}$ and $\boldsymbol{H}_{w32}$ are i.i.d. $\mathcal{CN}(0,1)$, and each block use of the channel corresponds to an independent realization of channel matrices.

\section{Capacity Bounds of MIMO Relay Channel}
In this section, we review the capacity upper bound and lower bound for MIMO relay channel, and present the results obtained for the case of only CSIR.
\subsection{Cut-Set (CS) Upper Bound}
Using the ``\emph{max-flow min-cut}'' theorem, Cover and El Gamal~\cite{CE} showed that the capacity of the full-duplex relay channel in terms of the channel mutual information is upper bounded by
\begin{equation}
\label{eq5}
C_{CS} = \max_{p(x_1,x_2)} \min{\{I(x_1;y_2,y_3|x_2),I(x_1,x_2;y_3)\}}
\end{equation}
where the maximization is with respect to the joint distribution of the source and relay signals. Considering ${\boldsymbol{x}_i\sim\mathcal{CN}(\boldsymbol{0},\boldsymbol{Q}_{ii}), i=1,2}$, where $\boldsymbol{Q}_{ii}$ is the covariance matrix of $\boldsymbol{x}_i$, the mutual information expressions in~\eqref{eq5} can be expressed for the MIMO relay channel as~\cite{WZH}
\begin{equation}
\label{eq6}
C_{CS} = \max_{\boldsymbol{Q}_{ii}:\mathrm{tr}(\boldsymbol{Q}_{ii})\leq P_i,~i=1,2} \min{(C_1,C_2)}
\end{equation}
\begin{equation}
\label{eq7}
C_1=\log \det(\boldsymbol{I}_N+\boldsymbol{H}_1\boldsymbol{Q}_{1|2}\boldsymbol{H}_1^H)
\end{equation}
\begin{equation}
\label{eq8}
C_2=\log \det(\boldsymbol{I}_{N_3}+\boldsymbol{H}_2\boldsymbol{Q}\boldsymbol{H}_2^H)
\end{equation}
where $\boldsymbol{H}_1=\begin{bmatrix}\boldsymbol{H}_{31}\\\boldsymbol{H}_{21}\end{bmatrix}$, $\boldsymbol{H}_2=\begin{bmatrix}\boldsymbol{H}_{31}&\boldsymbol{H}_{32}\end{bmatrix}$, $N=N_2+N_3$ and $\boldsymbol{Q}_{1|2}\triangleq\mathcal{E}[\boldsymbol{x}_1\boldsymbol{x}_1^{H}|\boldsymbol{x}_2]
=\boldsymbol{Q}_{11}-\boldsymbol{Q}_{12}\boldsymbol{Q}_{22}^{-1}\boldsymbol{Q}_{21}$ is the conditional covariance matrix and given by Schur complement of $\boldsymbol{Q}_{22}$ in $\boldsymbol{Q}$~\cite{BV}. The optimal distribution $p(x_1,x_2)$ in~\eqref{eq5} is Gaussian~\cite{CE}, and consequently the maximization of~\eqref{eq6} would be with respect to three covariance matrices $\boldsymbol{Q}_{11}$, $\boldsymbol{Q}_{22}$, and $\boldsymbol{Q}_{12}$.

When the channel matrices are random (due to fading) and the CSI is only known at the receivers, the optimal joint transmit covariance matrix $\boldsymbol{Q}$ in~\eqref{eq6} is diagonal. Using Jensen's inequality, the authors in~\cite{KGG2} showed that the equal power allocation is the optimal solution, i.e.,
\begin{equation}
\label{eq9}
\boldsymbol{Q}_{11}=\frac{P_1}{M_1}\boldsymbol{I}_{M_1},~~\boldsymbol{Q}_{22}=\frac{P_2}{M_2}\boldsymbol{I}_{M_2},~~
\boldsymbol{Q}_{12}=\boldsymbol{0}
\end{equation}
where $\boldsymbol{Q}_{12}=\boldsymbol{0}$ refers to the independence between the source and the relay signals. Thus, the CS upper bound for the MIMO relay channel with only CSIR can be expressed as
\begin{equation}
\label{eq10}
C_{CS}^{R}=\min(C_{1}^{R},C_2^R)
\end{equation}
\begin{equation}
\label{eq11}
C_1^R=\mathcal{E}\left[\log\det(\boldsymbol{I}_{N}+\frac{P_1}{M_1}.\boldsymbol{H}_1\boldsymbol{H}_1^H)\right]
\end{equation}
\begin{equation}
\label{eq12}
C_2^R=\mathcal{E}\left[\log\det(\boldsymbol{I}_{N_3}+\boldsymbol{H}_2\left[
\begin{matrix}
\frac{P_1}{M_1}.\boldsymbol{I}_{M_1} &  \boldsymbol{0}\\
\mathbf{0}                                            &  \frac{P_2}{M_2}.\boldsymbol{I}_{M_2}
\end{matrix}
\right]\boldsymbol{H}_2^H)\right]
\end{equation}
where the superscript $R$ stands for ``Rayleigh'', and the expectations are taken over the channel gains matrices.
\subsection{DF Achievable Rate }
The capacity of the full-duplex relay channel is lower bounded by the DF achievable rate~\cite{CE} given as
\begin{equation}
\label{eq13}
R_{DF} = \max_{p(\boldsymbol{x}_1,\boldsymbol{x}_2)} \min{\{I(\boldsymbol{x}_1;\boldsymbol{y}_2|\boldsymbol{x}_2),I(\boldsymbol{x}_1,\boldsymbol{x}_2;\boldsymbol{y}_3)\}}
\end{equation}
where the optimal distribution is again Gaussian and the maximization should be done over the joint distribution of the source and relay signals. In this strategy the relay first decodes the received signal from the source, and then re-encodes it before forwarding it to the destination. Considering~\eqref{eq4} and using the same approach for evaluation of the CS upper bound, it is easy to show that the capacity of MIMO relay channel with only CSIR is lower bounded by
\begin{equation}
\label{eq14}
R_{DF}^{R}=\min(C_{3}^{R},C_2^R)
\end{equation}
\begin{equation}
\label{eq15}
C_3^R=\mathcal{E}\left[\log\det(\boldsymbol{I}_{N_2}+\frac{P_1}{M_1}.b^2.\boldsymbol{H}_{w21}\boldsymbol{H}_{w21}^H)\right]
\end{equation}
where $C_2^R$ is the same as in~\eqref{eq12}. Note that~\eqref{eq15} is the ergodic capacity of MIMO channel between the source and the relay, and we will use it in the following section to find the optimal relay location in the sense of maximizing coverage region.
\section{Main Results}
In this section, considering a desired transmission rate, we define the concept of coverage in MIMO relay channel. Next, we investigate the coverage region and the optimal relay location based on: 1) an exact expression for the ergodic capacity of MIMO channels, and 2) an approximation in the high-SNR regime. Finally, we examine the impact of using multiple antennas on the optimal relay location.
\subsection{Coverage Definition}
Our definition of coverage region has a close relation to the concept of outage capacity~\cite{H}, and we consider it for MIMO relay channel as a geographic region at which a rate of at least $R>0$ is guaranteed, i.e.,
\begin{equation}
\label{eq16}
\mathcal{A}(u_2)\triangleq\{ d_3: C(u_2,d_3)\geq R\}
\end{equation}
where $R$ denotes the desired transmission rate in bps/Hz, $C(u_2,d_3 )$ is the channel capacity when there is a fixed distance $u_2$ between the source and the relay, and the destination is located at $d_3=(u_3,v_3)$.

Since the capacity of MIMO relay channel is still an open problem in general, it can be inferred from the condition in~\eqref{eq16} that the lower bound of the channel (DF achievable rate) should be larger than $R$. On the other hand, for applying DF strategy it is necessary for the relay to have the ability of decoding the information which is transmitted to it from the source. Thus, in order for the $R$ to be achievable by the source-relay channel, considering~\eqref{eq15}, the desired transmission rate $R$ can be defined as follows
\begin{equation}
\label{eq17}
R\triangleq\mathcal{E}\left[\log\det(\boldsymbol{I}_{N_2}+\frac{\rho}{M_1}\boldsymbol{H}_{w21}\boldsymbol{H}_{w21}^H)\right]
\end{equation}
where $\rho\triangleq P_1.{d^{*}}^{-\eta}$ is the effective SNR, and $d^{*}$ is a boundary distance which determines whether the DF strategy can be applied or not, since the condition $u_2\leq d^{*}$ guarantees that the relay is still able to decode the transmitted signal from the source. We define $d^{*}$ as the optimal relay location (for a desired transmission rate $R$) at which the coverage region is maximum. Our intuition for this finding is based on the fact that when the relay locates at distances smaller than $d^*$ (${u_2<d^*}$) the coverage region decreases, while for distances larger than $d^*$ ($u_2>d^*$) the DF strategy can not be used anymore in accordance with our definition of coverage in~\eqref{eq16}. Therefore, $u_2=d^*$ is the best choice for relay location in the sense of maximizing coverage.

\subsection{Desired Transmission Rate Analysis}
Note that~\eqref{eq17} involves the expectation operator that generally admits no explicit solution. In order to obtain a theoretical expression between the desired transmission rate $R$ and the optimal relay location $d^*$, in what follows, we evaluate $R$ by using two analytical approaches.
\subsubsection{Exact Expression for Desired Transmission Rate}
As stated before, the $R$ in~\eqref{eq17} is the capacity of source-relay MIMO channel, thus it can be expressed by the exact ergodic capacity of uncorrelated Rayleigh fading MIMO channel with only CSIR as [16, Theorem 9]
\begin{equation}
\label{eq18}
R=\frac{1}{\ln 2.\Gamma _m(m).\Gamma _m(n)}.\sum^m_{l=1}\det(\Psi (l))
\end{equation}
where $m=min(M_1,N_2)$, $n=max(M_1,N_2)$, and the modified multivariate Gamma function $\Gamma _t(k)$ and the auxiliary $m\times m$ matrix $\Psi (l)$ are defined as follows
\begin{equation}
\label{eq19}
\Gamma _t(k)\triangleq \prod^t_{p=1}\Gamma(k-p+1)
\end{equation}
\begin{equation}
\label{eq20}
\Psi(l)\triangleq \left[\left\{
\begin{matrix}
\Gamma(s)   &  i\neq l\\
\\
\Gamma(s).\mathcal{F}(1,\frac{\rho}{M_1},s)  &  i=l
\end{matrix}
\right]\right.
\end{equation}
where $1\leq i,j\leq m$, $s\triangleq n-m+i+j-1$, $\Gamma(k+1)=k!$, and the auxiliary function $\mathcal{F}(e,h,g)$ is defined by
\begin{figure}[!t]
	\centering
		\includegraphics[width=3.31 in]{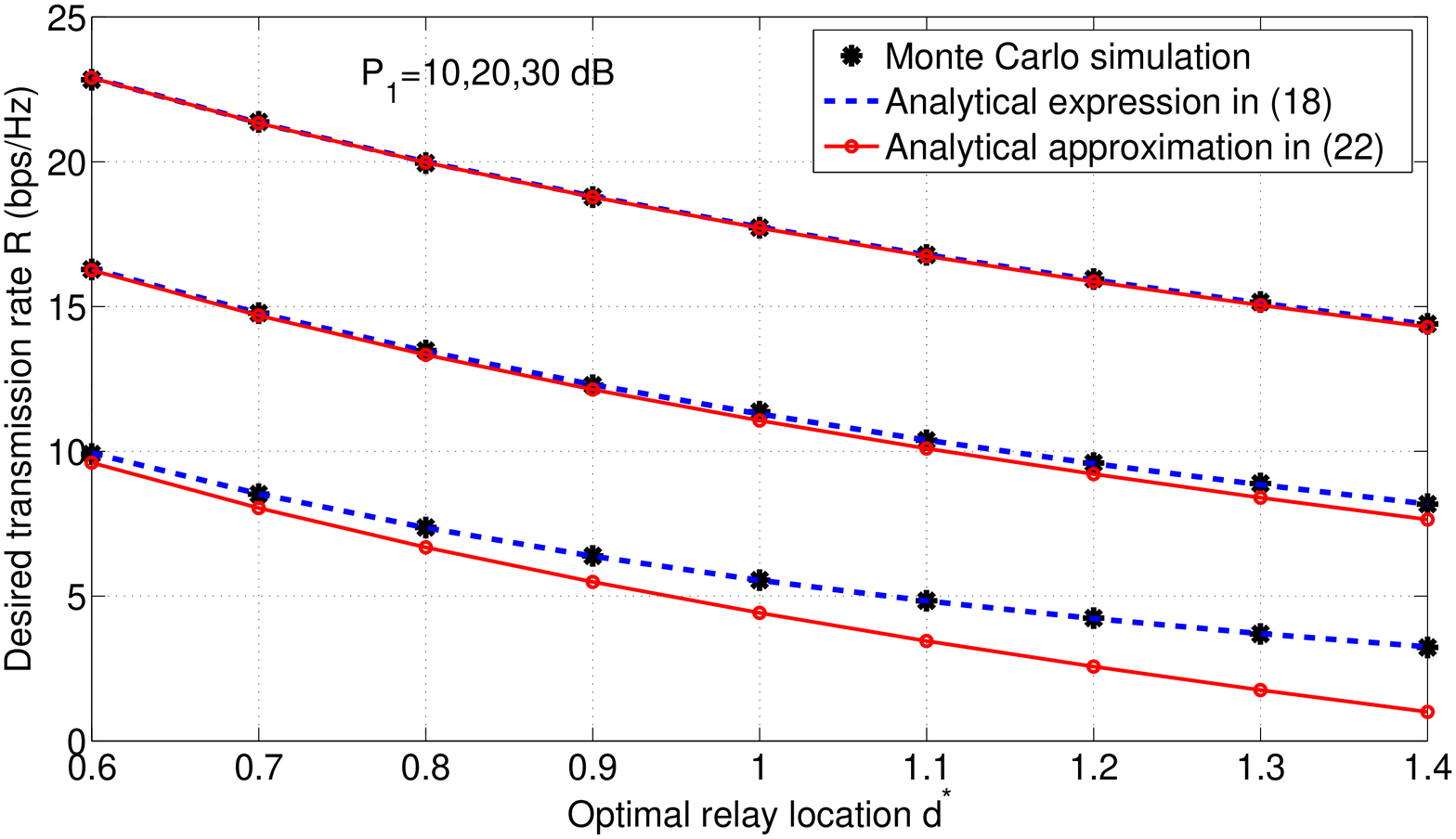}
		\caption{The relation between $R$ and $d^*$ when $M_1=N_2=2$.}
	\label{Fig3}
\end{figure}
\begin{equation}
\label{eq21}
\mathcal{F}(e,h,g)\triangleq h^{-g}.\exp^{\frac{1}{h.e}}.\sum^g_{p=1}(h.e)^p.\Gamma(-g+p ,\frac{1}{h.e})
\end{equation}
with incomplete Gamma function $\Gamma(-k,1/x)$ for positive integers $k$ defined as in [16, eq. (90)].

%\subsubsection{Approximation in the High-SNR Regime}
\subsubsection{1}

Now we restrict our attention to the high-SNR regime, and derive the following theorem.

\textit{Theorem 1}: In the high-SNR regime, the relation between the desired transmission rate $R$ and the optimal relay location $d^*$ can be approximated by the following expression
\begin{equation}
\label{eq22}
R\approx m.\log\left(\frac{\rho .\exp^{\psi(1)}}{M_1}\right)+\frac{1}{\ln 2}.\sum^m_{p=1}\sum^{n-p}_{q=1}\frac{1}{q}
\end{equation}
where $-\psi(1)\approx 0.577215$ is the Euler-Mascheroni constant.

\textit{Proof}:
For large $\rho$ we can ignore the identity matrix $\boldsymbol{I}_{N_2 }$ in~\eqref{eq17}, and simplify it as follows
\begin{equation}
\label{eq23}
R\approx \left\{
\begin{matrix}
\mathcal{E}\left[\log\det(\frac{\rho}{M_1}\boldsymbol{H}_{w21}\boldsymbol{H}_{w21}^H)\right],   &  M_1\geq N_2\\
\\
\mathcal{E}\left[\log\det(\frac{\rho}{M_1}\boldsymbol{H}_{w21}^H\boldsymbol{H}_{w21})\right],   &  M_1< N_2.
\end{matrix}
\right.
\end{equation}
Considering the case $M_1\geq N_2$, $R$ can be rewritten as
\begin{equation}
\label{eq24}
R\approx N_2.\log{\frac{\rho}{M_1}}+\frac{1}{\ln 2}.\mathcal{E}\left[\ln \det(\boldsymbol{H}_{w21}\boldsymbol{H}_{w21}^H)\right].
\end{equation}
Note that $\boldsymbol{H}_{w21}\boldsymbol{H}_{w21}^H$ is an i.i.d. Wishart matrix with $M_2$ degree of freedom and covariance matrix $\boldsymbol{I} _{N_2}$. Thus, the expectation in~\eqref{eq24} can be expressed as (see [17, Theorem 2.11])
\begin{equation}
\label{eq25}
\mathcal{E}\left[\ln \det(\boldsymbol{H}_{w21}\boldsymbol{H}_{w21}^H)\right]=\sum_{p=0}^{N_2-1}\psi(M_2-p)
\end{equation}
where $\psi(r)$ is Euler's digamma function, which for natural $r$ can be expressed as~\cite{GR}
\begin{equation}
\label{eq26}
\psi(r)=\psi(1)+\sum_{q=1}^{r-1}\frac{1}{q}.
\end{equation}
Thus,~\eqref{eq25} can be rewritten as
\begin{equation}
\label{eq27}
\mathcal{E}\left[\ln \det(\boldsymbol{H}_{w21}\boldsymbol{H}_{w21}^H)\right]=N_2.\psi(1)+\sum_{p=1}^{N_2}\sum_{q=1}^{M_2-p}\frac{1}{q}
\end{equation}
and applying~\eqref{eq27} to~\eqref{eq24}, the result follows. Note that by using the determinant identity $\det(I+AB)=\det⁡(I+BA)$ and following similar steps in~\eqref{eq24}-\eqref{eq27}, the desired transmission rate can be evaluated for the case $M_1<N_2.$
\begin{figure}[!t]
	\centering
		\includegraphics[width=3.5 in]{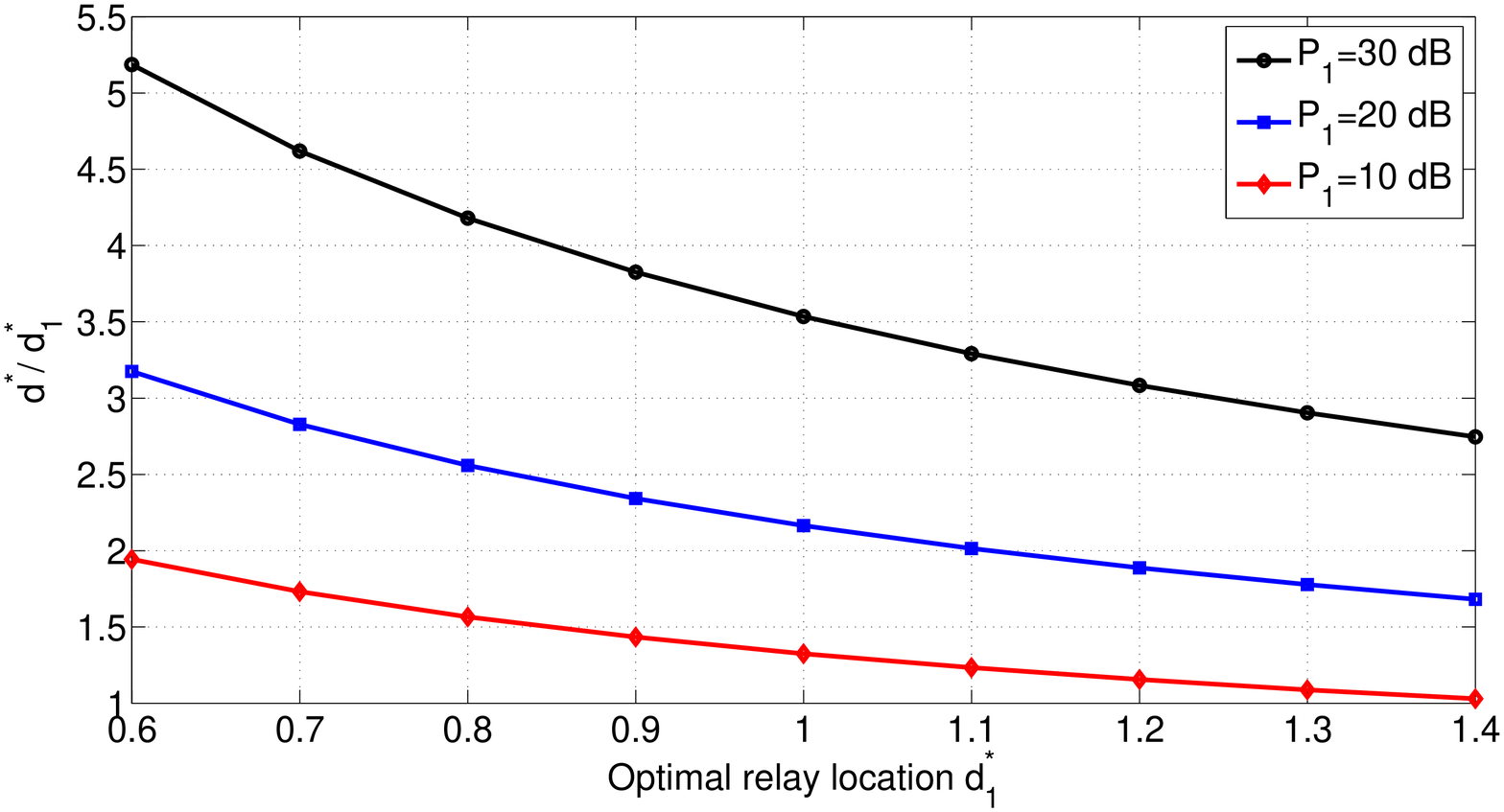}
		\caption{The relation between $d^*$ and $d_1^*$  when $R=R_1$, and $M_1=N_2=4$.}
	\label{Fig4}
\end{figure}
\subsection{The Impact of Using Multiple Antennas on the Coverage}
In this section, considering the high-SNR regime, we investigate the effect of using multiple antennas on the coverage region of MIMO relay channel in Rayleigh fading environment. Thus, we express $d^*$ in terms of the optimal relay location in single-antenna relay channel.

Using~\eqref{eq22}, the desired transmission rate for the single-antenna relay channel ($M_1=N_2=1$), can be written as
\begin{equation}
\label{eq28}
R_1\approx\log(P_1.{d_1^{*}}^{-\eta}.\exp^{\psi(1)})
\end{equation}
where $d_1^*$ denotes the optimal relay location in this channel. Considering~\eqref{eq22} and~\eqref{eq28} for a fixed transmission rate (${R=R_1}$), $d^*$ can be expressed as follows
\begin{equation}
\label{eq29}
d^*=\left(\frac{(P_1.\exp^{\psi(1)})^{m-1}.d_1^{*^{\eta}}.2^A}{M_1^m}\right)^\frac{1}{m.\eta}
\end{equation}
where $A$ is defined as
\begin{equation}
\label{eq30}
A\triangleq\frac{1}{\ln 2}.\sum_{p=1}^{m}\sum_{q=1}^{n-p}\frac{1}{q}.
\end{equation}

\textit{Remark}: Note that $d^*$ in~\eqref{eq29} has an inverse relationship with the number of transmit antennas, and thus we can conclude that receiver diversity provides wider coverage region than transmitter diversity.

\section{Numerical Results}
In this section we present the numerical results for the coverage region of MIMO relay channel. In our simulations we assume that $\eta=3.52$, ${M_1=N_2=M_2=N_3=N}$, and $P_1=P_2=10$~dB. Fig.~\ref{Fig3} depicts the desired transmission rate $R$ for different values of $d^*$. The region below each curves represents the region where DF strategy can be applied, while the region above contains the points in which this strategy can not be used anymore. It can be seen from the figure that there is a perfect match between Monte Carlo simulation and the analytical expression in~\eqref{eq18}, and also the approximation in~\eqref{eq22} is more accurate at high-SNR than at low-SNR.
\begin{figure}[t]
	\centering
		\includegraphics[width=3.5in]{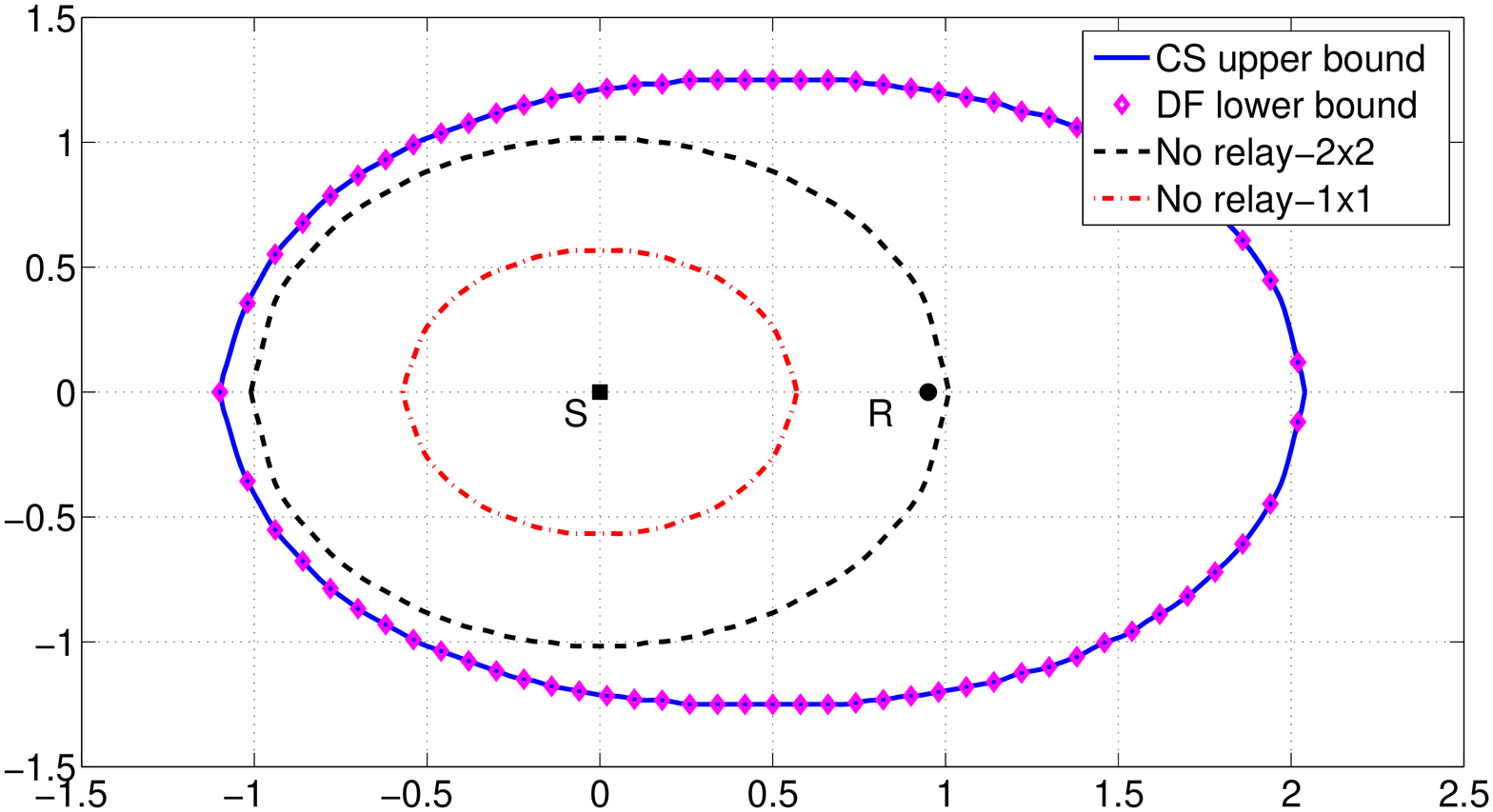}
		\caption{Coverage region when $u_2=0.95$, and $N=2$.}
	\label{Fig5}
\end{figure}
\begin{figure}[!t]
	\centering
		\includegraphics[width=3.5 in]{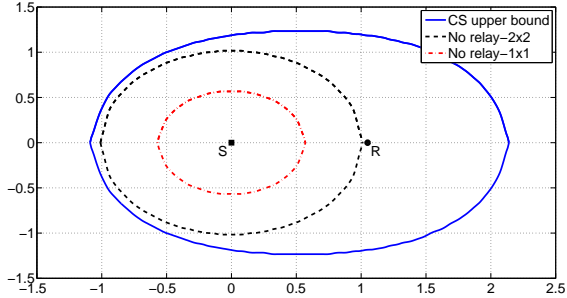}
		\caption{Coverage region when $u_2=1.05$, and $N=2$.}
	\label{Fig6}
\end{figure}

In Fig.~\ref{Fig4} the effect of using multiple antennas on the optimal relay location is investigated. This figure shows the ratio $d^*/{d_1^*}$ for different values of transmit power versus the optimal relay location for single-antenna relay channel. It can be inferred from the figure that as the distance between the source and the relay gets larger, the effect of using multiple antennas decreases.

For desired transmission rate $R=5.5$, the optimal relay location is $d^*=1$ [see \eqref{eq18}]. Fig.~\ref{Fig5} shows the coverage region for $u_2=0.95$ while in Fig.~\ref{Fig6} it is assumed that $u_2=1.05$. For performance comparison, these figures also include the coverage region of the ``no-relay'' cases. Fig.~\ref{Fig5} indicates that if $u_2<d^*$ the capacity bounds converge but the coverage decreases. On the other hand, if $u_2>d^*$, it can be inferred form the Fig.~\ref{Fig6} that although the CS upper bound may extend, but in this case the DF strategy can not be used anymore.

\begin{figure}[t]
	\centering
		\includegraphics[width=3.5 in]{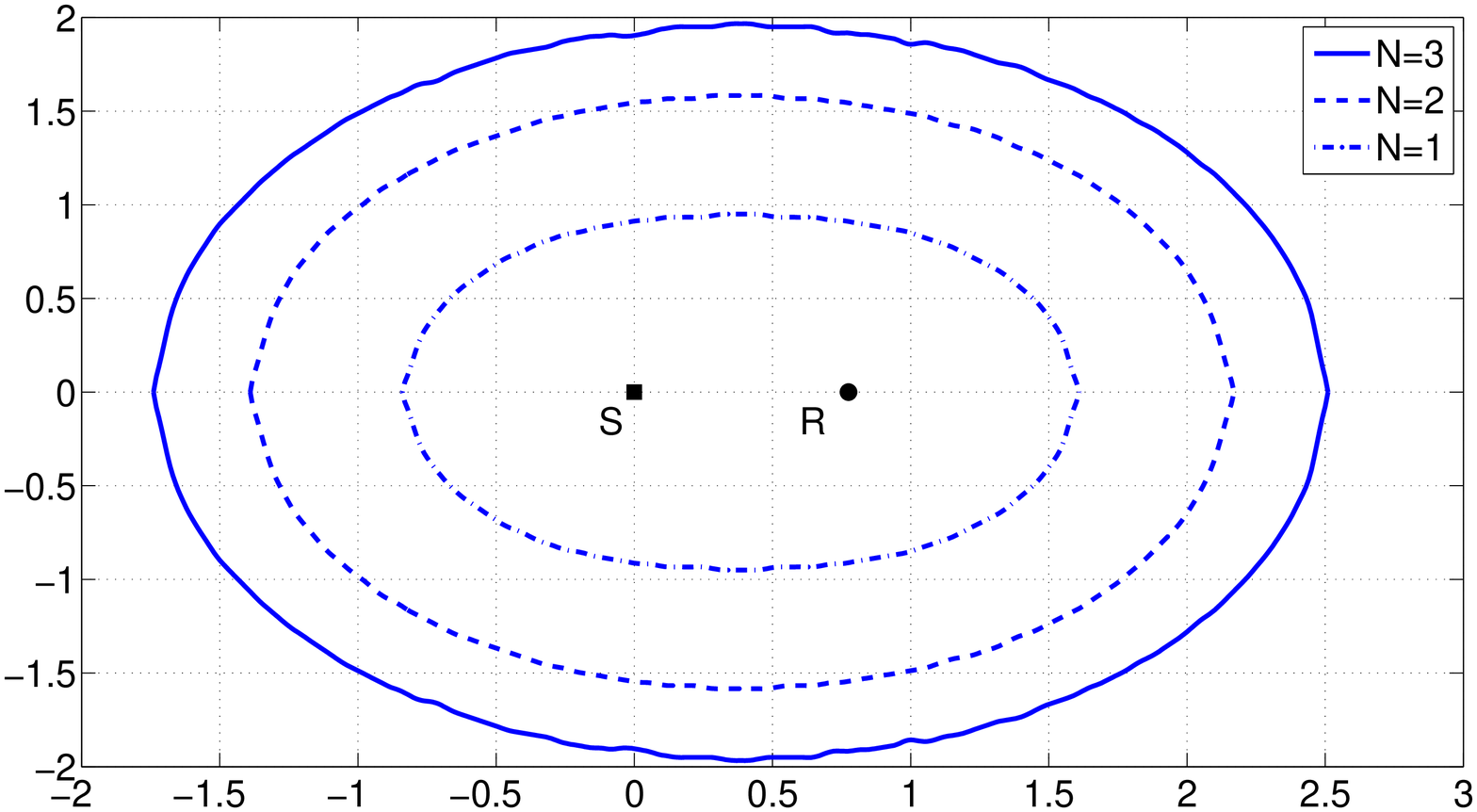}
		\caption{Coverage region when $R=4$, and $u_2=0.775$.}
	\label{Fig7}
\end{figure}

\begin{figure}[!t]
	\centering
		\includegraphics[width=3.38 in]{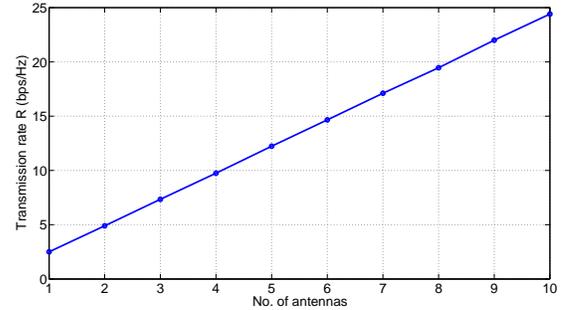}
		\caption{Transmission rate vs. number of antennas for a fixed coverage.}
	\label{Fig8}
\end{figure}

Fig.~\ref{Fig7} shows the growth of coverage region for a fixed transmission rate with increasing the number of antennas, and Fig.~\ref{Fig8} depicts the growth of transmission rate for a fixed coverage. It can be observed that for a fixed coverage region there is a linear relationship between the transmission rate and the number of antennas.
\section{Conclusion}
In this paper we investigated the optimal relay location in the sense of maximizing coverage region for MIMO relay channel. Considering the case when the channel matrix entries are i.i.d. Rayleigh fading and the CSI is only available at the receivers, we defined the coverage region for MIMO relay channel, and derived an exact analytic expression with the help of which we could determine the optimal relay location (for a desired transmission rate) at which the coverage region is maximum. Also, an approximation is presented for the high-SNR regime. Numerical results confirm the accuracy of our analysis, and show that using multiple antennas increases coverage region for a fixed transmission rate, and also increases the transmission rate almost linearly for a fixed coverage region.
%***********************************************************************

\end{document}